\documentclass[conference]{IEEEtran}
\IEEEoverridecommandlockouts
\usepackage{cite}
\usepackage{amsmath,amssymb,amsfonts}
\usepackage[pdftex]{graphicx}
\usepackage{textcomp}
\usepackage{xcolor}

\usepackage{caption}
\usepackage{algorithm}
\usepackage{algpseudocode}
\usepackage{subcaption}
\usepackage{bm}
\usepackage{mathtools}
\usepackage{amssymb}
\usepackage{multirow}
\usepackage{mathrsfs}
\usepackage{soul}
\usepackage{xcolor}
\usepackage{adjustbox}
\usepackage{listings}
\usepackage{xcolor}

\definecolor{codegreen}{rgb}{0,0.6,0}
\definecolor{codegray}{rgb}{0.5,0.5,0.5}
\definecolor{codepurple}{rgb}{0.58,0,0.82}

\definecolor{backcolour}{rgb}{0.95,0.95,0.92}

\definecolor{blue}{RGB}{0, 0, 255}      
\definecolor{red}{RGB}{255, 0, 0}       
\definecolor{green}{RGB}{0, 255, 0}     

\lstdefinestyle{json}{
    backgroundcolor=\color{backcolour}, 
    basicstyle=\small\ttfamily,
    breaklines=true,
    frame=single,
    rulecolor=\color{black},
    showstringspaces=false,
    numbers=left,
    numberstyle=\tiny\color{gray},
    numbersep=5pt,
    morekeywords={assert, property, Prompt, Response}, 
    keywordstyle=\color{red}, 
}

\definecolor{codegreen}{rgb}{0,0.6,0}

\lstdefinestyle{myStyle}{
    backgroundcolor=\color{backcolour},   
    commentstyle=\color{codegreen},
    basicstyle=\ttfamily\footnotesize,
    breakatwhitespace=false,         
    breaklines=true,                 
    keepspaces=true,      
    frame=single,           
    numbers=left,       
    numbersep=5pt,                  
    showspaces=false,                
    showstringspaces=false,
    showtabs=false,                  
    tabsize=2,
}

\newcommand*{\ToolName}{AssertCraft}
\begin{document}

\title{\huge{
Automatic High-quality Verilog Assertion Generation through Subtask-Focused Fine-Tuned LLMs and Iterative Prompting  		\vspace{-0.2in}
} }

\author{Mohammad Shahidzadeh\textsuperscript{1}, Behnam Ghavami\textsuperscript{2}, Steve Wilton\textsuperscript{2}, Lesley Shannon\textsuperscript{1}
\\
\textsuperscript{1}Simon~Fraser~University, Canada\\
\textsuperscript{2}University of British Columbia, Canada\\
}




\maketitle
\vspace{-0.5in}

\begin{abstract}
\small
Formal Property Verification (FPV), using SystemVerilog Assertions (SVA), is crucial for ensuring the completeness of design with respect to the specification. However, writing SVA is a laborious task and has a steep learning curve. In this work, we present a large language model (LLM) -based flow to automatically generate high-quality SVA from the design specification documents, named \ToolName. 
We introduce a novel sub-task-focused fine-tuning approach that effectively addresses functionally incorrect assertions produced by baseline LLMs, leading to a remarkable 7.3-fold increase in the number of functionally correct assertions. Recognizing the prevalence of syntax and semantic errors, we also developed an iterative refinement method that enhances the LLM's initial outputs by systematically re-prompting it to correct identified issues. This process is further strengthened by a custom compiler that generates meaningful error messages, guiding the LLM towards improved accuracy. The experiments demonstrate a 26\% increase in the number of assertions free from syntax errors using this approach, showcasing its potential to streamline the FPV process.

\end{abstract}


\section{Introduction}\label{sec:intro}
The slowdown in processing power following Moore's law has introduced a significant rise in hardware diversity, resulting in an increased need for design verification (DV)  \cite{orenes2023autocc}. Originally, DV was performed by extracting the properties of the design and extensively trying to prove them by finding corner cases and new simulation testbenches. However, in the earlier cycles of DV, this tedious and uncertain task was replaced by formal verification \cite{6469140}. The formal method replaced this uncertain task with the automatic process of proving that the property is correct under all circumstances or there is a counter-example proving otherwise. The properties (assertion statements) in the design are written in Property Specification Language (PSL), containing an assert keyword and a property inside them which are used to check the correctness of that property. These properties come from \textit{design specifications} and show the expected functionality of the Design Under Test (DUT). However, each design has a separate specification and infinite ways of implementation. As a result, the formal verification engineers need to generate independent assertions by reading the \textit{design specification} and the \textit{RTL design}, which makes this process tedious and time-consuming. This hurdle prevented engineers from using it more commonly \cite{seligman2023formal}.  Given the recent increasing complexity of designs, the demand for more efficient methods to streamline this process has become critical.

A number of studies have attempted to automate parts of the assertion extraction process. Earlier efforts in this domain sought to simplify this task by providing a new abstraction level that is closer to human language \cite{orenes2021autosva} \cite{huang2018instruction}. However, full automation has not been achieved, as the specification must still be converted to this new abstraction level rather than directly to the assertion level. 
lately, large language models (LLM) have demonstrated significant promise in automating hardware design and assisting engineers throughout the design process \cite{wang2024software}. This has led to promising developments in using LLMs to automate the verification process \cite{kande2023llm,kande2024security,orenes2023using}. Kande et al. \cite{kande2023llm} illustrated how LLMs can generate new assertions from detailed comments, while Orenes et al. \cite{orenes2023using} showcased their effectiveness in creating new properties from the Register Transfer Level (RTL) code. Despite these advancements, no successful attempt has yet been made to generate high-quality assertions directly from \textit{specification documents}.



Our research, detailed in Section \ref{sec:challenges}, investigates the challenges faced by current state-of-the-art large language models in generating assertions directly from specification documents. This process requires complex, multi-step reasoning, which is a known challenge for Pretrained LLMs (PLLMs), as demonstrated in recent studies on other applications \cite{talmor2020olmpics} \cite{kassner2020pretrained} \cite{rae2021scaling}. 
As a promising solution, fine-tuning LLMs is essential and widely adopted to unlock robust capabilities for complex tasks, enhance performance on downstream applications, and better align with human preferences \cite{min2019multi}. However, this approach relies on the availability of \textbf{"large-scale, annotated task-specific training data"} accumulated over time. This dependency limits the practical use of LLMs-based assertion generation models in industrial verification scenarios where such training data is scarce. While fine-tuning the LLM for assertion generation from specifications may present significant challenges, creating a dataset for a subtask of the original problem is feasible. Therefore, we introduce a novel \textit{“sub-task-focused fine-tuning method,”} which involves dividing the task into subtasks that can be fine-tuned separately.  While the method may yield sensible initial outputs, some of the assertions generated using this method can still contain compilation bugs that must be addressed. 
However, having enough assertions without compilation bugs is essential to hit a high level of coverage and enter the late cycles of the design \cite{hu1997formal}\cite{kern1999formal}. So this challenge should be addressed efficiently.
To tackle these issues, we introduce a \textit{"iterative bug-fixing paradigm"} that tries to build upon its previous patch by using a refinement loop between the compiler and the debugger LLM. This paradigm, unlike the comment one-step paradigms \cite{xia2022less} \cite{zhu2021syntax} would not miss the multi-location bugs and is good at catching small bugs \cite{ye2024iter}. 
In this work, for the first time, we brought this multi-step bug-fixing paradigm into the assertion generation flow by adding a customized compiler that can generate meaningful refinement prompts for the LLM. Alongside, to assess the proposed flow's ability to generate high-quality assertions, we developed a comprehensive evaluation tool-set that includes a dataset and a scoreboard for measuring assertion quality. Our experiments revealed an impressive improvement of 7.3 times in the number of functionally correct assertions compared to the baseline PLLM, i.e. GPT-3.5. Additionally, we achieved near-perfect (100\%) stimuli coverage on several designs.

 \begin{figure*}[!t]
 	\begin{center}
 	    \vspace{-0.2in}
 		\includegraphics[width = 1\textwidth]{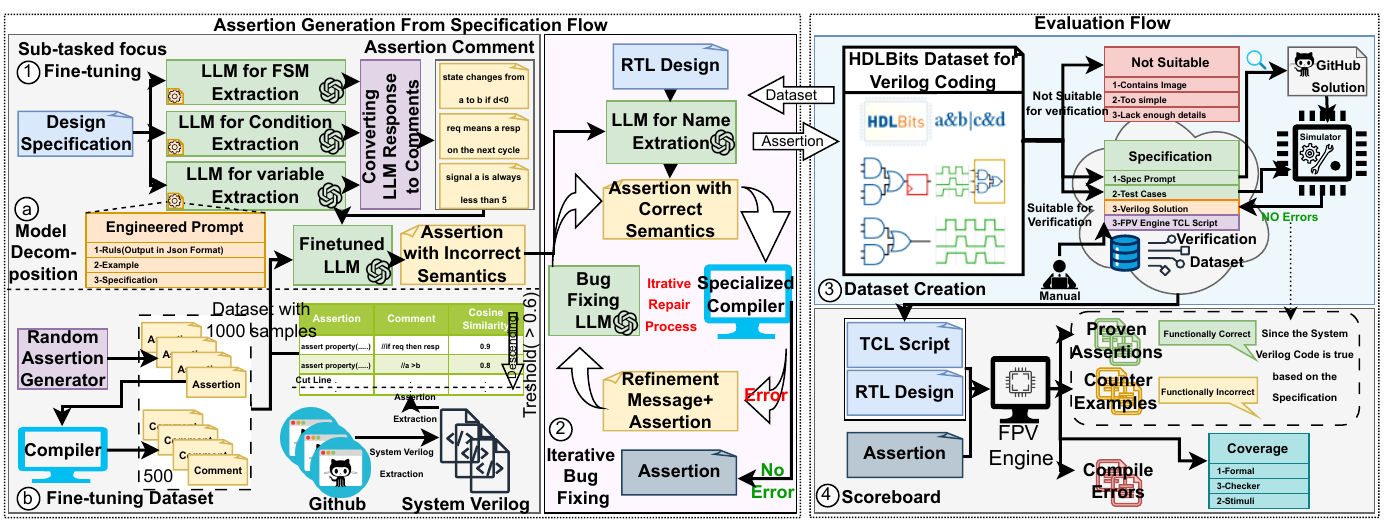}
 		\vspace{-0.2in}
 		\caption{The proposed \ToolName{} full-stack flow employs two novel techniques—sub-tasked fine-tuning, and iterative prompting—to produce high-accuracy assertion statements from the specification document. Moreover, this flow was completed by adding an additional dataset and scoreboard for assessing the ability of the model to generate high-quality assertions. }
 		\vspace{-0.1in}
 		\label{fig:Flow}
 	\end{center}	
 \vspace{-0.2in}
 \end{figure*}

The key contributions of this paper are summarized as follows:

\begin{itemize}
    \item Launch of a fully automated toolset for RTL assertion generation from design specifications.
    \item Introduction of a novel sub-tasked focused fine-tuning method for assertion generation which improves the ability of GPT to produce functionally correction assertions by 7.3x.
    \item Development of an Iterative repair method for assertion generation using a specialized compiler engineered to produce meaningful error messages for LLMs, which is able to fix 26\% assertions with bugs.
     \item Creation of a dataset for evaluating LLM-based assertion generation, which includes specifications and reference code to establish a robust benchmark for assessing the quality of generated assertions.
\end{itemize}

\section{Related Work and Motivation}

\label{sec:related}

\subsection{Automation in Assertion Generation}

Earlier works in automatic assertion generation, such as AutoSVA \cite{orenes2021autosva} and ILA \cite{huang2018instruction}, have focused on creating a new abstraction level closer to human language. Although they succeeded in making the assertion generation process easier, they lost their generality and did not entirely solve the process, as engineers still had to develop new properties at these abstraction levels.

Another endeavor in assertion automation targets the common patterns in the waveform for assertion generation \cite{vasudevan2010goldmine}. It first uses data mining techniques to discover patterns in simulation test benches, then formulates assertions to describe these patterns. Although this work achieves full automation in assertion generation, it suffers from a high number of generated assertions and the possibility of the assertions being wrong due to the waveform being generated using a design under test, which might have numerous bugs. In response, Liu et al. \cite{liu2011automatic} and \cite{liu2012word} sought to address these issues by utilizing higher abstraction levels for waveform generation, such as transaction-level modeling, and by enhancing their techniques for ranking generated assertions, thereby reducing their overall number. Despite these improvements, their approaches still fail to extract functionally correct assertions from specification documents and rely on additional assumptions, such as access to simulation traces or a higher abstraction level, rendering them semi-automated. Additionally, Witharana et al. \cite{witharana2022survey} provide a comprehensive survey of recent advancements in automated techniques for hardware verification using simulation traces.

With the rise of new LLMs \cite{thakur2024verigen,GitHubCopilot,pearce2020dave} capable of accurately generating simple RTL code similar to humans, recent research has shifted focus to the use of LLMs for verification. Kande et al. \cite{kande2023llm,kande2024security} addressed a fundamental research question: Can LLMs be used to generate assertions from comments? To explore this, they generated assertions at three levels of detail and assessed whether the LLM-generated assertions were accurate based on the provided comments. Their findings suggest that LLMs can produce correct assertions from comments, but only when the comments are detailed and include all relevant signal names and operations. The latest work in this area \cite{orenes2023using} demonstrated the capability of LLMs to generate assertions from RTL code by incorporating additional rules into the LLM's prompt. Although both works were successful in generating correct assertions, they made some impractical assumptions. For example, \cite{kande2023llm} uses a well-defined comment to produce assertions instead of extracting them from the specification, and \cite{orenes2023using} may generate assertions from buggy RTL code, which can lead to incorrect assertions.

\subsection{Challenges in PLLM-Based Assertion Generation}
\label{sec:challenges}
LLMs like GPT-2 and BERT are transformer-based artificial neural networks designed to operate on text datasets. These models contain millions to billions of parameters and are trained on vast amounts of text data. Both the inputs and outputs of an LLM consist of tokens, which are common sequences of characters.  When given a sequence of tokens as a prompt, an LLM generates a probability distribution for the next token based on its vocabulary. Once a token is chosen according to specific criteria, it is added to the prompt, and the LLM proceeds to generate the next token in a process known as auto-regression. 

\begin{figure}[!htb]
 	\begin{center}
 	    \vspace{-0.02in}
 		\includegraphics[width = 0.5\textwidth]{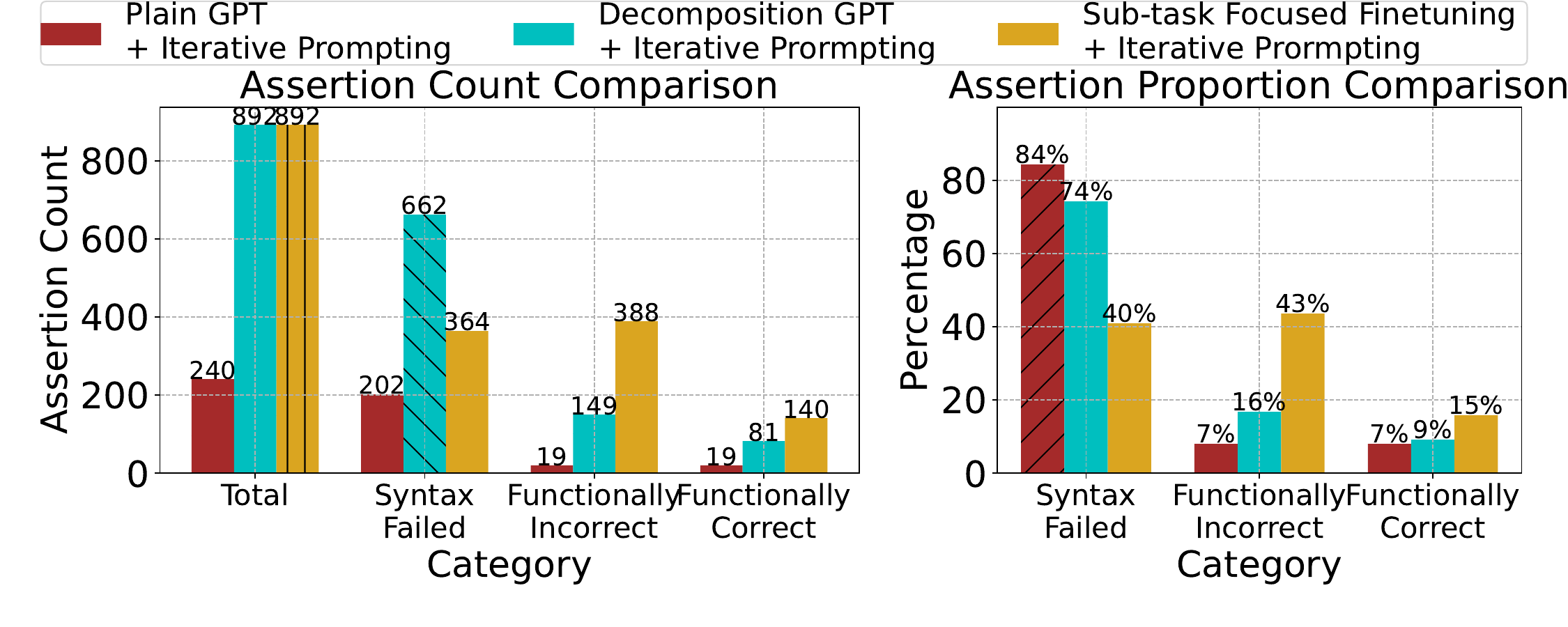}
 		\vspace{-0.2in}
 		\caption{Comparison of assertion generation across three scenarios: employing Plain GPT alone, modifying Plain GPT with comments extracted via the first phase of Sub-task Focused Fine-tuning, and \ToolName{} which includes a custom GPT for assertion generation and comment extraction. All these models also include the iterative prompting method to fix syntax and semantic errors.}
 		\vspace{-0.1in}
 		\label{fig:comaprision_gpt}
 	\end{center}	
 \vspace{-0.2in}
 \end{figure}

One way to generate assertions from the specification is to prompt the PLLM to create assertions without additional fine-tuning or details. We attempted to generate assertions directly from the specification document using PLLM, and the results are illustrated in Figure \ref{fig:comaprision_gpt}. As shown, only 240 assertions were generated for all the modules in the dataset, out of a potential 892, with only 19 of those being functionally correct. These poor results stem from the complexity of generating assertions from the specification. Listing 1 further illustrates why so few assertions were produced using this naive approach. In this example, we presented the PLLM with a simple Finite State Machine specification and asked it to generate assertions. However, instead of producing multiple assertions for each transition between states, it attempted to encapsulate the entire concept in a single assertion. As a result, the generated assertion was not only syntactically incorrect due to multiple implied statements, but also functionally incorrect according to the specification. In fact, PLLMs can only generate accurate assertions when features such as signal names and their relationships are described with precision.

\section{Proposed \ToolName}
\label{sec:proposed}
The overall proposed flow, named as \ToolName{}, is illustrated in Figure \ref{fig:Flow}. The automation process begins with a clear English specification outlining the RTL code's functionality. This specification must be comprehensive enough for the engineer to write the RTL code based solely on the provided information. Once the specification and the corresponding RTL design are established, the proposed automation flow verifies the design's implementation through the generated assertions. Essentially, the tool autonomously generates assertions and enhances their quality without requiring external input. The figure illustrates the assertion generation and evaluation flows of \ToolName{}, presented separately on the left and right sides. In the assertion generation section, it accepts the design specification and RTL design as input and produces assertions. \ToolName{} generates these assertions in two steps. In step (1), \ToolName{} employs a focused fine-tuning method to generate an initial assertion from the design specification. This assertion then undergoes an iterative repair process, which aims to correct any syntax and semantic errors, marked as step (2) in the flow. On the right side of the figure, the evaluation flow is presented to assess the quality of the generated assertions. To achieve this, It first builds a novel dataset for this task in step (3). Ultimately, it evaluates the quality of the results on the dataset using a scoring system.

\subsection{Subtask-Focused Fine-Tuning}
\label{sec:information}

\ToolName's assertion generation process from the design specification and the RTL code has two main steps, as depicted in Figure \ref{fig:Flow}. In the first step, Subtask-Focused fine-tuning, only the design specification is given to the tool for the initial assertion generation, and the tool is expected to generate assertions.

\subsubsection{LLM}
We employed GPT-3.5-turbo\cite{chatgpt_ref} as the Language Model for its fine-tuning capabilities and cost-effectiveness compared to GPT-4 \cite{gpt4_ref}. Additionally, its utilization is free if the API is not accessed. 
We used the default fine-tuning setting recommended by the OpenAI and fine-tuned the model for 3 epochs.




\subsubsection{Model Composition}

We propose to enhance the generation of assertions from specifications using LLMs by fine-tuning the model. However, we find this process challenging because a dataset containing both specifications and assertions does not exist, and generating one would consume an enormous amount of time and resources. Nevertheless, a robust dataset containing comments for the assertions and the assertions themselves can be created by scraping GitHub repositories that contain SystemVerilog files. As a result, we divided the task into two \textit{"sub-tasks"} which can be individually improved using fine-tuning or zero-shot prompting techniques. First, we generate comments from the specification. Every individual comment is then used to make the assertions using a fine-tuned LLM which is trained for assertion generation from the comments.


We use three questions to extract meaningful comments from the specification. Listings 2, 3, and 4 outline the inquiries used to break the specification into smaller components. The first question in Listing 2 is directed at FSMs in the design. For every FSM, we extract the states and the conditions to transfer from one state to another. This helps to break each FSM into multiple assertions, each indicating one transition between the states. Question B in Listing 3 was tailored to extract conditional statements, and Question C to extract the variable ranges. Each of these pieces of information was then treated as one assertion in the remainder of the flow.

\subsubsection{Fine-Tuning}
To fine-tune GPT, we first created a robust dataset containing comments and their corresponding assertions. Figure \ref{fig:Flow}--part (b) illustrates the creation process of this dataset. Two approaches were pursued to make this dataset. Initially, existing assertions within GitHub repositories were extracted to be used in the dataset. However, only some of the extracted assertions had well-defined comments, necessitating the process of data cleaning and selecting the best matching comments and assertions. Consequently, a sorting mechanism based on the cosine similarity of their word2vec vectors was implemented, setting a threshold from 0.6 onwards to filter out irrelevant assertions, as seen in the right section of Figure \ref{fig:Flow}--step(1)-part (b).
Ultimately, this process yielded 500 relevant examples. Additionally, another 500 assertions were incorporated, randomly generated by an automated script, which were then matched with predefined sentences for each operation to form the comment for the assertion. Our dataset served as the foundation for refining GPT3.5 Turbo over three epochs. Finally, we modified the system message for better results by integrating content from a SystemVerilog property specification cheat sheet \cite{wang_svassertions}.



\vspace{-0.1in}
\begin{lstlisting}[style=json, caption={Plain GPT Example}, label={lst:firstq1}]
Prompt:
"The code receives a stream of data then a 4 
-stage FSM detects the 1101 pattern
and outputs 1 wherever it sees this pattern."
+rules from [9]. 
---------------------------------------------
Response:
assert property (|-> (name.pattern_detected |=> name.output_wire));
assert property (name.pattern_detected_reg |=> name.output_wire_reg);
\end{lstlisting}

\vspace{-0.18in}

\begin{lstlisting}[
    caption=Question A,
    label={lst:firstq},
    language=C++,
    backgroundcolor=\color{backcolour},   
    commentstyle=\color{black},
    keywordstyle=\color{magenta},
    numberstyle=\tiny\color{codegray},
    stringstyle=\color{codegreen},
    basicstyle=\ttfamily\footnotesize,
    breakatwhitespace=false,         
    breaklines=true,                 
    keepspaces=true,                 
    numbers=left,       
    numbersep=5pt,                  
    showspaces=false,                
    showstringspaces=false,
    showtabs=false,                  
    tabsize=2
    ]
/*specify the following on the provided text and return the answer in the JSON format:
1- states in the FSM
2-conditions to move from one state to another in the FSM(plus the state when the condition is false. empty if the state is not determined in the text)
3- conditions for the output of the FSM and the state which produces this output.
Note:(Do not make any additional states based on your assumption)
Example JSON:
keep the structure only for the field names and change the values*/
[{
  "states": [""],
  "transitions": [
    {
      "current_state": "",
      "conditions": "",
      "next_state_condition_true": "",
      "next_state_condition_false": "",
    }],
  "outputs":[ {
      "current_state": "",
      "output_name": "",
      "conditions": "",
      "output_value_condition_true": "",
      "output_value_condition_false": "",
    }],
  }
},..,number of fsms] 
\end{lstlisting}

\begin{lstlisting}[
    caption=Question B,
    label={lst:secondq},
    language=C++,
    backgroundcolor=\color{backcolour},   
    commentstyle=\color{black},
    keywordstyle=\color{magenta},
    numberstyle=\tiny\color{codegray},
    stringstyle=\color{codegreen},
    basicstyle=\ttfamily\footnotesize,
    breakatwhitespace=false,         
    breaklines=true,                 
    keepspaces=true,                 
    numbers=left,       
    numbersep=5pt,                  
    showspaces=false,                
    showstringspaces=false,
    showtabs=false,                  
    tabsize=2
    ]
/*specify every conditional statement on the provided text and return the answer in the JSON format:

Note:(Do not make any additional assumptions)
Example JSON:
keep the structure only for the field names and change the values */
[{
  "antecedent": "",
  "consequent": ""
    }],
  }
},.., number of conditions] 
\end{lstlisting}



\vspace{-0.18in}

\begin{lstlisting}[
    caption=Question C,
    label={lst:thirdq},
    language=C++,
    backgroundcolor=\color{backcolour},   
    commentstyle=\color{black},
    keywordstyle=\color{magenta},
    numberstyle=\tiny\color{codegray},
    stringstyle=\color{codegreen},
    basicstyle=\ttfamily\footnotesize,
    breakatwhitespace=false,         
    breaklines=true,                 
    keepspaces=true,                 
    numbers=left,       
    numbersep=5pt,                  
    showspaces=false,                
    showstringspaces=false,
    showtabs=false,                  
    tabsize=2
    ]
/*Specify every variable in the text, indicate the range it can take under different conditions, and return the answer in JSON format. If the information is not provided for a field, leave it empty.

Example JSON:
Keep the structure only for the field names and change the values:*/
[{
  "variable_name": "",
  "condition_list": [
    {"condition": "",
     "range_or_value": ""}
  ]
}]
\end{lstlisting}
 		\vspace{-0.05in}

Figure \ref{fig:comaprision_gpt} demonstrates the impact of custom GPT on result quality. Notably, it reduces assertions with failed syntax errors by 33\% while nearly doubling the number of functionally correct assertions.
\vspace{-0.05in}
\subsection{Iterative Repair}
Using a sub-task fine-tuned model significantly improved the quality of generated assertions, the assertions generated in this step had only access to specification. This prevents the model from generating incorrect assertions due to buggy RTL code but increases the risk of semantic mistakes since the model does not know about the implementation and the signal names. In our experiment, shown in Table \ref{tab:assertions}, we found that 78\% of the assertions were failing due to syntax errors before reaching this step. As a result, we incorporated an iterative repair method to address this issue. This process is illustrated in Figure \ref{fig:Flow}—step (2). First, the RTL code and the assertions are provided to an LLM, which is tasked with correcting any semantic errors. Next, the assertions are submitted to a compiler for correctness verification. If the generated assertion is correct, it is output as the final assertion for the model. However, if the assertion is incorrect, the compilation error is sent back to the LLM for correction. This process is repeated until we obtain an assertion free of any syntax or semantic errors, or until we reach a predefined iteration threshold where a correct assertion cannot be generated. Our findings indicate that in such cases, the model may become stuck in an indefinite loop, repeatedly providing the same answer.

Through our investigation, we found that GPT struggles with tasks that require step-by-step thinking. For example, it cannot detect the simple task of finding the $i^{th}$ word in the text, which is the most commonly used error message for humans. As a result, the first change in our custom compiler was the annotation of the error part without merely indicating the location of the error. The second change involved a simple addition for combinational circuits. Since our model was struggling to generate correct assertions for combinational circuits, we tasked it with converting the sequential assertion into a combinational assertion by adding a set of rules. At this stage, we instructed the LLM to delete all clock-related functions and sections in the assertion and convert the implications to the combinational mode. For example, if the compiler encountered the keywords ($a|->b$ or $a=>b$), we tasked it to convert them to ($!a | b$). 



\begin{table}[tb!]
\centering
\caption{Summary of Assertion Counts}
 \vspace{-0.1in}
\label{tab:assertions}
\begin{tabular}{|l|l|}
\hline
\textbf{Title}                                & \textbf{Number of Assertions} \\ \hline
Generated Assertions                          & 892                            \\ \hline
Assertions With Incorrect Syntax             & 364                            \\ \hline
Assertions With Correct Syntax               & 528                            \\ \hline
Assertions with Fixed Error                  & 234                            \\ \hline
Assertions without errors                    & 294                            \\ \hline
\end{tabular}
 \vspace{-0.2in}
\end{table}


\vspace{-0.1in}
\subsection{Evaluation Flow}
Evaluating any LLM-based assertion generation process requires a structured flow due to the complexity and variability of the generated assertions, which complicate consistent accuracy assessment. To measure \ToolName{}'s effectiveness, we created an evaluation flow that includes a dataset and a scoreboard. This general approach can be applied to evaluate any LLM-based assertion generation process.

\subsubsection{Dataset Creation}
\label{sec:Dataset}
 This dataset helps us measure the quality of generated assertions by providing a TCL script, Specification, and the correct RTL design of that specification.
The dataset in Figure \ref{fig:Flow}--part(3) is derived from a subset of the HDLBits dataset \cite{hdlbits}, which includes specifications for various hardware designs. Several enhancements were made to adapt this dataset for verification purposes. First, the golden RTL code was added to the specification for each design from a GitHub repository \cite{hdlbits_solutions}. Although these RTL codes were functionally correct in accordance with the simulation result from the HDLBits website, they were not all synthesizable. As a result, the first subtle modification was making the designs synthesizable and compatible with FPV engines. Subsequently, a TCL script was written for them to prepare it for the FPV engine execution. Table \ref{tab:hdl_bits} enumerates all the Verilog designs within the HDLBits dataset. However, not all of the designs were suitable for verification tasks, as some had images in their specifications or lacked descriptive specifications from which to make assertions. The right-hand side of the table showcases the designs that were used in our dataset. Throughout this paper, these modules are referenced using a hierarchical numbering scheme (e.g., 1.2.0.3), which signifies the first entry of category 1, the second entry of category 2 (excluding category 3), and the third utilized module (in this case, "Replicate").

\begin{table*}[tb!]
\centering

\vspace{-0.4in}
\caption{Comparison of modules between the HDL-Bits dataset \cite{hdlbits} and the new dataset introduced in this paper. The new dataset comprises modules listed within the 'Used Modules' section of HDL-Bits, covering various categories.}
\label{tab:hdl_bits}

\scalebox{0.85}{
\begin{tabular}{|p{1.5cm}|p{1.5cm}|p{1.6cm}|p{8.2cm}|p{7.8cm}|}
\hline
\textbf{Category-1} & \textbf{Category-2} & \textbf{Category-3} & \textbf{Not Used Modules} & \textbf{Used Modules} \\
\hline
\multirow{5}{*}{1-Verilog} & 1-Basics &  & Simple Four wires, Inverter, AND, NOR, XNOR, Declare wires, 7458 chip & \\
\cline{2-5}
& 2-Vectors  &  & Vectors, Vectors (detail),   Four-input gates & 1-Vector part select, 2-Bitwise ops, 3-Replicate, 4-More replication, 5-reversal 1, 6-Vector concat\\
\cline{2-5}
  Language & 3-Module Hierarchy &  & Modules, Connect ports by position, Connect ports by name, Three modules, Modules and vectors, Adder 1, Adder 2,
Carry-select, Adder-subtractor & \\
\cline{2-5}
 & 4-Procedures  &  & Always blocks (combinational), Always blocks (clocked), If statement, Case statement, Priority encoder,
Priority encoder with casez & 1-Avoiding latches, 2-If statement latches\\
\cline{2-5}
 & 5-More Features  & & Reduction operators, Reduction: Wider gates,  Combination
for-loop: 255-bit count, Generate for-loop: 100-bit adder 2, Generate for-loop: 100-digit BCD adder & 1-Combination for-loop: Vector reversal 2, 2-Conditional ternary\\
\hline
\multirow{10}{*}{2-Circuits} & \multirow{5}{*}{1-Comb.}& 1-Basic Gates & Wire, GND, NOR, Another, Two gates, More gates, 7420 chip, Simple circuits A, B, Combine
circuits A, B, 3-bit count & 1-Truth tables, 2-Gates and vectors, 3-Two-bit equality,  4-longer vectors,  5-Ring or vibrate?, 6-Thermostat\\
\cline{3-5}
 &  & 2-Multiplexer  & 2-to-1, 2-to-1 bus mux, &  1-(256-to-1 4-bit), 2-(256-to-1), 3-(9-to-1)\\
\cline{3-5}
 &  Circuits & 3-Arithmetic Circuits  & Half add, Full add, 3-bit adder, 100-bit binary adder, 4-digit BCD adder & 1-Signed addition overflow,\\
\cline{3-5}
 &  & 4-K-maps &  &  1-K-map with a mux, 2-Minimum SOP and POS, 3-3 variable, 4-4-variable 2 5-4 variable 3 6-karnaugh map 1, 7-karnaugh map 2, 8-(4-variable 1)\\
\cline{2-5}
 & \multirow{5}{*}{2-Seq. } & 1-Latches and FFs & DFFs, DFF (reset), DFF (reset value), DFF (asynch.), DFF (byte enable), D Latch, DFF, DFF+gate, Mux and DFF, DFFs and
gates &  1-Circuit from truth table, 2-Edge capture register, 3-Detect edge/both edges, 4-Dual-edge triggered FF\\
\cline{3-5}
 &  & 2-Counters & Counter 1000, 4-digit
decimal counter &  1-Slow decade counter,2-Decade counter again, 3-Four-bit binary counter, 4-12-hour clock,  5-Counter 1-12, 6-Decade counter\\
\cline{3-5}
 & Circuits & 3-Shift Registers & 5-bit/3-bit/LSFR& 1-Left/right arithmetic shift by 1/8, 2-4-bit shift register, 3-Left/right rotate, 4-3-input LUT , 5-Shift register, 6-32-bit LFSR\\
\cline{3-5}
 &  & 4-Cellular Automata & & 1-Rule 90, 2-Rule 110, 3-Conways Game of Life 16x16\\
\cline{3-5}
 &  & 5-FSM & FSM 1 (asynch.), FSM 1 (synch.), FSM 2 (asynch.), FSM 2 (synch.), Simple state transitions 3, Simple one-hot state transition 3, Sequence recognition, Q5b: Serial twos complementer (Mealy FSM), Q2a, Q6b: FSM next-state logic, Q6c: FSM one-hot next-state logic, Q6: FSM, Q2a: FSM, Q2b: One-hot FSM &  1-PS/2 packet parser and datapath, 2-Serial receiver with parity check, 3-Q2b, 4-Q3b: FSM, 5-Q8: Design Mealy FSM, 6-Q3c: FSM
logic, 7-Q5a: Serial twos complementer (Moore FSM), 8-FSM 3 (asynch.), 9-Moore FSM, 10-PS/2 packet parser,  11-One-hot FSM, 12-Q3a, 13-Serial receiver, 14-Serial receiver and datapath, 15-FSM 3 (synch.), 16-Q3b\\
\cline{3-5}
 &  & 6-Larger Circuits & FSM: One-hot logic &  1-Counter with period 1000, 2-FSM: Sequence 1101 recognizer,  3-4-bit shift register and down counter, 4-Complete timer, 5-FSM: Complete FSM, 6-FSM: Enable shift register\\
\hline
3-Fix Bugs  &  &  & Mux2, NAND, Mux4, Add/subtract, Case statement & \\
\hline
4-Write Test &  & & Clock, T flip-flop & \\
\hline
\end{tabular}
}

\vspace{-0.3in}
\end{table*}

\subsubsection{Scoreboard}

The final step in the evaluation platform is a pass to the FPV engine. This step determines the quality of results for the entire flow. The FPV engine categorizes the assertions into three main categories. The first type is the syntactically incorrect assertions. These assertions can have incorrect syntax, like a missing parenthesis, or be semantically incorrect, like incorrect naming for signals. The second category consists of the assertions with the correct syntax for which the FPV engine can find a counterexample. Since the RTL code is the golden RTL and it is in accordance with the specification, these assertions are labeled as functionally incorrect. Finally, the assertions with no syntax error or counterexample within the 20 cycles of the FPV run were labeled as functionally correct. We chose 20 cycles as the threshold since it was sufficiently accurate in its prediction.

\section{Experimental Results}
\label{sec:results}
In this section, we share the results from the evaluation flow on different modules in table \ref{tab:hdl_bits}. In the following sections, we showed the ability of \ToolName{} to generate functionally correct assertions and assess the coverage of generated assertions.


\subsection{Correctness}
\begin{figure}[!htb]
 	\begin{center}
 		\includegraphics[width = 0.5\textwidth]{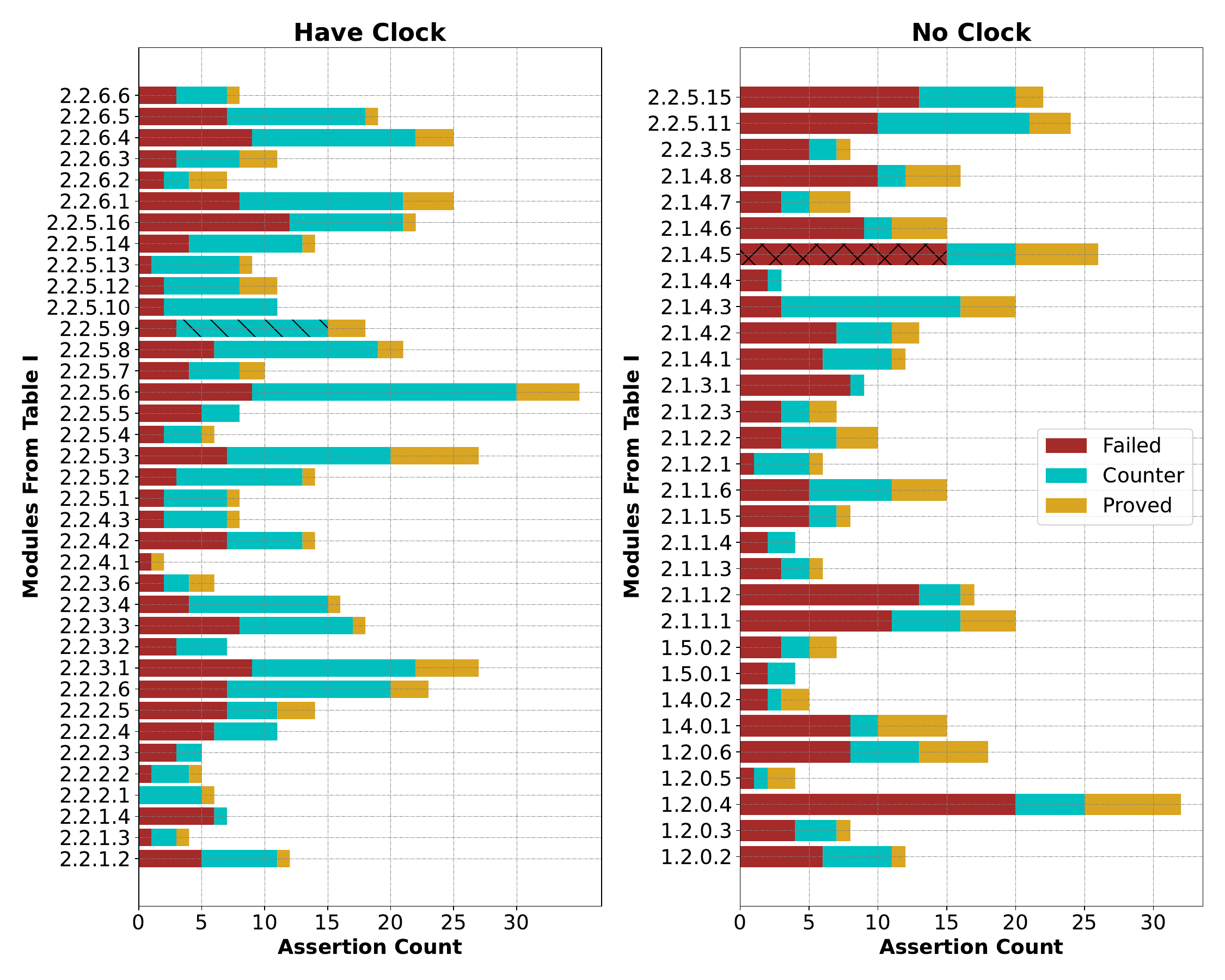}
 		\caption{The quality of the generated assertions for each design in the HDLBits dataset \cite{hdlbits} is represented in a bar chart. The red bar indicates the number of assertions with syntax or semantic errors. The blue bar illustrates the number of functionally incorrect assertions. Finally, the yellow bar shows the number of functionally correct assertions.}
 		\label{fig:test_results}
 	\end{center}	
 \vspace{-0.4in}
 \end{figure}

Figure \ref{fig:test_results} provides a more detailed overview of the generated assertions for each Verilog module in the dataset in table \ref{tab:hdl_bits}. The y-axis illustrates the filename using the indexing format introduced in section \ref{sec:Dataset}, and the x-axis shows the number of assertions that are functionally correct (yellow bar), functionally incorrect (blue bar), and syntactically or semantically incorrect (red bar). The figure is divided into two general sections: modules with clock and reset signals and modules without clock or reset signals. This distinction is important in assertion generation as the patterns for combinational and sequential assertions are completely different.

Our analysis reveals that in most modules, \ToolName{} was able to generate at least one functionally correct assertion. Additionally, the number of failed assertions due to incorrect syntax is considerably higher in modules without a clock, as it is more challenging to generate these assertions for such modules compared to those with concurrent assertions.


\subsection{Coverage}
\begin{figure}[!htb]
 	\begin{center}
 		\includegraphics[width = 0.5\textwidth]{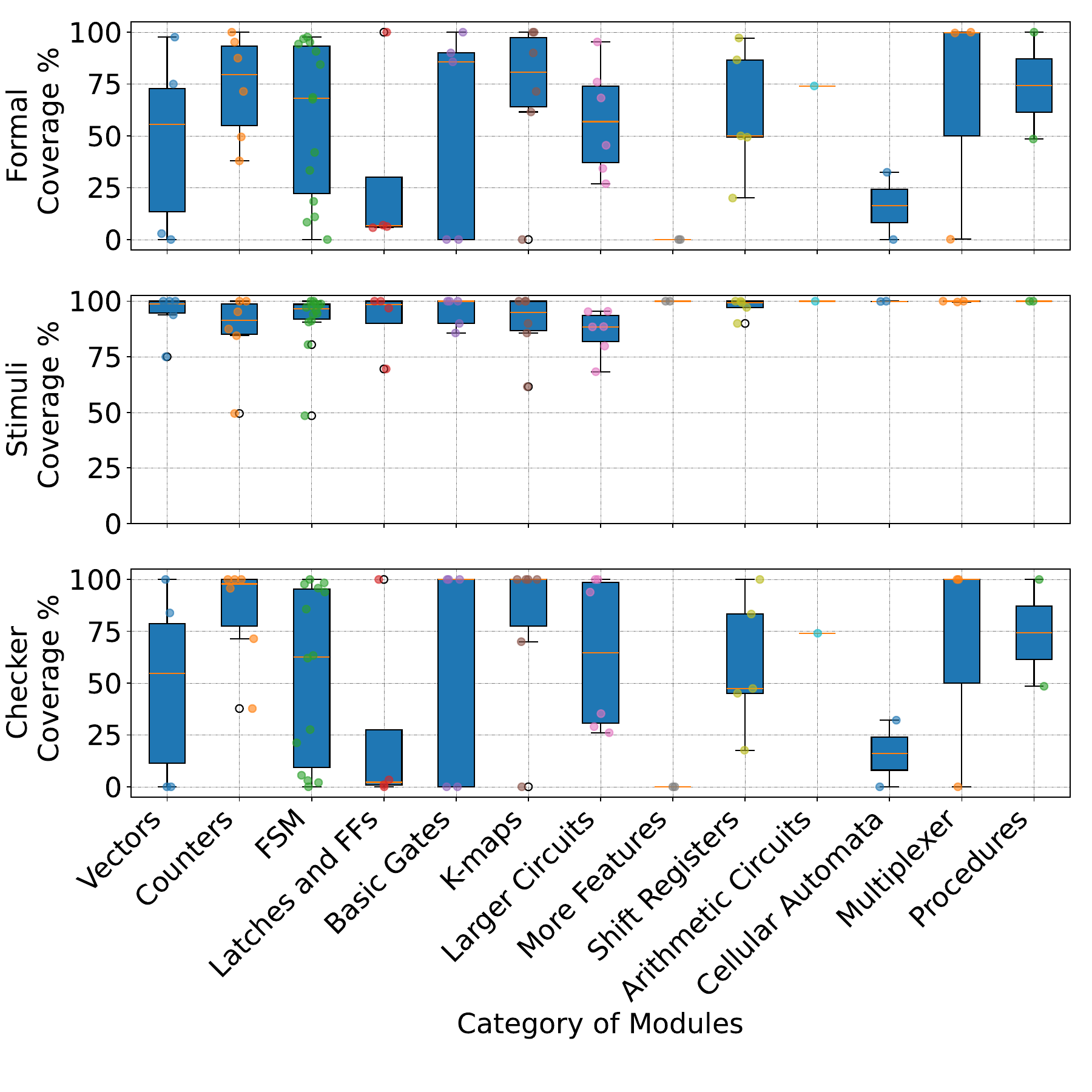}
 		\vspace{-0.3in}
 		\caption{Coverage results for each HDLBits category \cite{hdlbits} are shown as distributions. Coverage is represented as a range of values due to multiple designs, with individual data points also included.
   }
 		\vspace{-0.1in}
 		\label{fig:Coverage}
 	\end{center}	
 \vspace{-0.3in}
 \end{figure}

We computed the coverage for each module in the dataset and aggregated the results for each category of modules. Coverage is one of the fundamental metrics in formal verification which is used as an indicator to move to the next stages of the design.  Figure \ref{fig:Coverage} illustrates the coverage distribution for each category of data. As can be seen, the stimuli coverage is almost always between 80\% and 100\% for all categories, while the checker and formal coverages vary significantly based on the category. However, the mean of the distribution is more than 50\% in most scenarios.

\section{Conclusion}
\label{sec:concl}
In this paper, we showcased the current state of pretrained LLMs in the assertion generation process and identified the challenges faced by state-of-the-art LLMs in this complex task. We also introduced a novel sub-task focused fine-tuning method that can improve the accuracy of LLMs by 7.3 times on this task. Furthermore, we refined the final output of the model using an iterative refinement method which can direct the LLM toward assertions without syntax or semantic errors, which leads to a 26\% increase in the number of correct assertions. While our framework demonstrates the potential of LLMs in generating high-quality assertions, further exploration is needed to fully leverage their capabilities in verification tasks. For instance, we could apply LLMs to automated theorem proving, where they can assist in formulating logical proofs. Additionally, exploring their application in formal methods for model checking could also yield insights into verifying system properties more efficiently. 




\section{Acknowledgements}
 We acknowledge the support from the Natural Sciences and Engineering Research Council (NSERC) of Canada, This work is funded by the Natural Sciences and Engineering Research Council of Canada NSERC https://www.nserccrsng.gc.ca/ under Grant No. NETGP485577-15 NSERC (COHESA project) and 341516 NSERC (RGPIN), along with in-kind support from AMD and Intel/Altera.
\bibliographystyle{IEEEtran}
\small
\bibliography{references}	
\end{document}